\documentstyle[twoside,fleqn,espcrc2,epsfig]{article}

\newcommand{\AmS}{{\protect\the\textfont2
  A\kern-.1667em\lower.5ex\hbox{M}\kern-.125emS}}

\begin{document}
\title{Anderson localization in Hubbard ladders}

\author{ E. Orignac and T. Giamarchi
\address{Laboratoire de Physique
Des Solides, U.P.S. B\^at 510, 91405 Orsay, France}
\thanks{Laboratoire associ{\'e} au CNRS. Email:
orignac@lps.u-psud.fr, giam@lps.u-psud.fr}
}
\begin{abstract}
The effect of a weak random potential on  two-leg Hubbard ladders is investigated.
The random potential is shown to induce Anderson localization except for attractive
enough interactions, supressing completely d-wave superconductivity.
These localization effects remain very strong even for many ladders
coupled by Josephson coupling. Both dc and ac conductivities
and localization lengths are obtained. Consequences for the
superconducting ladder compound Sr$_x$Ca$_{14-x}$Cu$_{24}$O$_{41+\delta}$ are
discussed.
\end{abstract}
\maketitle

The two-chain Hubbard ladder is a toy model of a metal with a
spin-gap, reminiscent of the metallic phase of underdoped cuprates.
Recent experiments\cite{uchara_SrCaCuO} on the superconducting transition in
Sr$_{14-x}$Ca$_x$Cu$_{24}$O$_{41}$ have been interpreted in the
framework of this model. However, impurities are only considered as a
source of holes and not a source of random potential although it is
known that an infinitesimal disorder  localizes all electronic states in one
dimension. It is therefore important to investigate the disordered
Hubbard ladder.
The disordered Hubbard ladder is defined by the Hamiltonian:
\begin{eqnarray}\label{definition}
H&=& -t_\parallel \sum_{i,p,\sigma}
(c^\dagger_{i+1,p,\sigma}c_{i,p,\sigma}+c^\dagger_{i,p,\sigma}
c_{i+1,p,\sigma})\nonumber \\
&-& t_\perp
\sum_{i,\sigma}(c^\dagger_{i,1,\sigma}c_{i,2,\sigma}+c^\dagger_{i,2,\sigma}c_{i,1,\sigma})\nonumber \\
&+&U\sum_{i,p}n_{i,p,\uparrow}n_{i,p,\downarrow} + \sum_{i,p}\epsilon_{i,p}n_{i,p}
\end{eqnarray}
Where $\epsilon$ is a random potential with
$\overline{\epsilon_{i,p}\epsilon_{j,q}}=D\delta_{i,j}\delta_{p,q}$.
For zero disorder $(D=0)$,
bosonization techniques allow to reduce
the Hamiltonian
(\ref{definition}) to a Hamiltonian with two bosonic charge modes and
two bosonic spin modes. Among those, only the mode corresponding to
total charge excitations is gapless. The ladder is thus in a metallic
state with a spin gap $\Delta$. The total charge
mode is completely characterized by
the velocity of charge excitations $u_{\rho+}$ and an
exponent $K_{\rho+}$. for attractive interactions, $K_{\rho+}>1$ and
for repulsive interactions $K_{\rho+}<1$. With attractive
interactions, s-wave superconducting fluctuations are dominant,
whereas for repulsive interactions,  d-wave superconducting
fluctuations are dominant \cite{schulz_2chains}.

The effect of a weak disorder $D\ll \Delta$ can be
considered in a Renormalization Group
analysis \cite{orignac_2chain_long}. For $K_{\rho+}<3/2$,
the disorder is relevant and the system
is in an Anderson localized phase. In particular, the d-wave
superconducting ``phase'' of the ladder is unstable with respect to
infinitesimal disorder, and the s-wave superconducting ``phase'' is
stable only for attractive enough interactions.

\begin{figure}[htbp]
  \begin{center}
    \epsfig{file=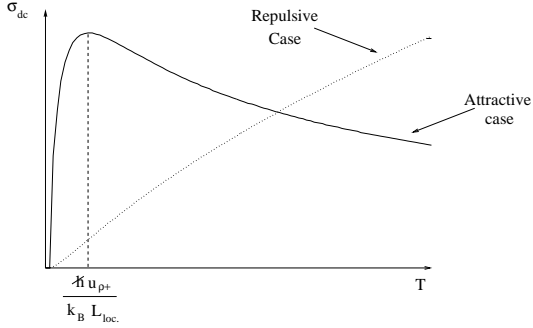,angle=-90,width=7cm}
    \caption{The behavior of d.c. conductivity as a function of temperature
      for repulsive and for attractive interactions.}
    \label{figure_dc}
  \end{center}
\end{figure}

\begin{figure}[htbp]
  \begin{center}
    \epsfig{file=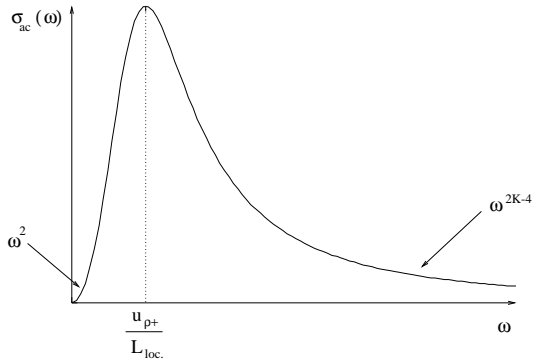,angle=-90,width=7cm}
    \caption{The behavior of a.c. conductivity (for $T=0$) as a function
      of frequency in the localized regime $K=K_{\rho+}<3/2$.}
    \label{figure_ac}
  \end{center}
\end{figure}

The localization length $L_{loc.}$ and both the dc and ac transport can
also be computed using the same RG procedure \cite{orignac_2chain_long}
and are shown respectively on Fig.~\ref{figure_dc} and Fig.~\ref{figure_ac}.
The temperature dependent conductivity is
\begin{equation}\label{cond_temp}
\sigma_{dc}(T)\sim T^{2-2K_{\rho+}}
\end{equation}
in the regime $k_BT \gg \frac{\hbar u_{\rho+}}{L_{loc.}}$.
In the localized regime and for
$k_BT\ll \frac{\hbar u_{\rho+}}{L_{loc.}}$, the conductivity
behaves as $\sigma_{dc}(T)\sim \exp \left(-(T_0/T)^{1/2}\right)$.

As can be
seen from Fig.\ref{figure_dc} and (\ref{cond_temp})
there is a maximum in $\sigma_{dc}$ as a
function of $T$ for attractive interactions which occurs
when the thermal coherence
length is of the order of magnitude of the localization length
$T\sim \frac{u_{\rho+}}{k_B L_{loc.}}$.  This maximum
is therefore a remnant of the s-wave superconductivity in the pure
system. Conversely,
there is \emph{no remnant} of the d-wave superconductivity of
the pure system for repulsive interactions. Although the
pure system has dominant d-wave superconductive fluctuations
the transport properties in presence of disorder
ressemble the one of an insulator.
These predictions on transport could be tested in ladder compounds
but also in quantum wire systems.

In the single ladder, d-wave superconductivity
resulting from
purely repulsive interactions is unstable in the presence of
infinitesimal disorder. To determine if this result persists
in a system of coupled ladders we have performed a mean field
treatment of Josephson coupled
ladders \cite{orignac_2chain_long}. Such a system would have a genuine
ordered d-wave phase and a finite crticial temperature in the pure case.
Even in the presence of interladder
Josephson coupling d-wave
superconductivity is very unstable with respect to non magnetic
disorder, leading again to a destruction of d-wave superconductivity for realistic
disorder strength. Let us note that this istability is
much stronger than a simple pair-breaking effect, and is due here
to the localization effect coming from the \emph{one-dimensional}
nature of the ladder \cite{orignac_2chain_long}.

Both the temperature dependence of the conductivity of a purely
one-dimenensional ladder model and the extreme sensitivity of $T_c$ to
impurities even for coupled ladders, strongly suggests
\cite{orignac_2chain_long} that the superconducting
transition \cite{uchara_SrCaCuO} in
Sr$_x$Ca$_{14-x}$Cu$_{24}$O$_{41+\delta}$
cannot be explained in terms of
stabilization of one-dimensional d-wave fluctuations by interchain
coupling and that a better theory of superconductivity in these
compounds should start from a truly two-dimensional limit.

\end{document}